\documentclass[table,xcdraw]{article}
\usepackage{spconfnew,amsmath,graphicx}

\usepackage{flushend} 
\usepackage[citecolor=black,
            colorlinks=true,
            linkcolor=black,
            urlcolor  = black,
            pdfpagelabels={true},
            pdftitle={On Time Domain Conformer Models for Monaural Speech Separation in Noisy Reverberant Acoustic Environments},
            pdfauthor={William Ravenscroft, Stefan Goetze, and Thomas Hain},
            pdfkeywords={speech separation, conformer, speech enhancement, time domain, single channel},
            pdfsubject={speech separation, conformer, speech enhancement, time domain, single channel}]
            {hyperref}
\usepackage[english]{babel}
\addto\extrasenglish{}
\addto\extrasenglish{}
\addto\extrasenglish{}
\addto\extrasenglish{}
\addto\extrasenglish{}
\addto\extrasenglish{}

\usepackage{todonotes}
\usepackage{acronym} 
\usepackage{svg}
\usepackage{url}
\usepackage{orcidlink}
\usepackage{censor}
\usepackage{amsfonts}
\usepackage{amsmath}
\usepackage{amssymb}
\usepackage{xcolor}
\usepackage{hhline}

\acrodef{AC}    {acoustic conditions}
\acrodef{ASR}   {automatic speech recognition}
\acrodef{BLSTM} {bidirectional long short term memory}
\acrodef{CB}    {convolutional block}
\acrodefplural{CB}[CBs]{convolutional blocks}
\acrodef{Conv-TasNet} {convolutional \ac{TasNet}}
\acrodef{conformer}{convolution augmented transformer}
\acrodef{CSM}   {clean speech mixtures}
\acrodef{cLN}   {channelwise layer normalization}
\acrodef{DAE}   {denoising autoencoder}
\acrodef{DE}    {density estimation}
\acrodef{D-Conv}{depthwise convolution}
\acrodef{DD-Conv}{deformable depthwise convolution}
\acrodef{DDS-Conv}{deformable depthwise-separable convolution}
\acrodef{DS-Conv}{depthwise-separable convolution}
\acrodef{DM}    {dynamic mixing}
\acrodef{DL}    {deep learning}
\acrodef{DNN}   {deep neural network}
\acrodef{DP}    {dual-path}
\acrodef{DPRNN} {dual-path recurrent neural network model}
\acrodef{DPTNet}{dual-path transformer network}
\acrodef{DTCN}  {deformable temporal convolutional network}
\acrodef{E2E}   {end-to-end}
\acrodef{ER}    {early reflection}
\acrodef{FFT}   {fast Fourier transform}
\acrodef{gLN}   {global layer normalization}
\acrodef{GLU}   {gated linear unit}
\acrodef{GRU}   {gated recurrent unit}
\acrodef{GC}    {global context}
\acrodef{FLOP/s} {floating point operations per second}
\acrodef{LSTM}  {long short term memory}
\acrodef{LR}    {late reflection}
\acrodefplural{LR}[LRs]{late reflections}
\acrodef{LUFS}   {loudness units relative to full scale}
\acrodef{MACs}  {mutiply-acccumulate operations}
\acrodef{MHA}   {multihead attention}
\acrodef{MHSA}  {multihead self-attention}
\acrodef{MOS}   {Mean Opinion Score}
\acrodef{MR}    {mask refinement}
\acrodef{NMF}   {non-negative matrix factorization}
\acrodef{NSM}   {noisy speech mixture}
\acrodef{NRSM}  {noisy reverberant speech mixture}
\acrodef{OOM}   {out-of-memory}
\acrodef{PE}    {positional encoding}
\acrodef{PESQ}  {perceptual evaluation of speech quality}
\acrodef{PIT}   {permutation invariant training}
\acrodef{PM}    {post-masking}
\acrodef{P-Conv}{pointwise convolution}
\acrodef{PReLU} {parametric \ac{ReLU}}
\acrodef{Q1}    {1st quartile}
\acrodef{Q2}    {2nd quartile}
\acrodef{Q3}    {3rd quartile}
\acrodef{Q4}    {4th quartile}
\acrodef{QDP}   {quasi-dual-path}
\acrodef{QDPN}  {quasi-dual-path network}
\acrodef{ReLU}  {rectified linear unit}
\acrodef{RF}    {receptive field}
\acrodef{RIR}   {room impulse response}
\acrodef{RNN}   {recurrent neural network}
\acrodef{RSM}   {reverberant speech mixture}
\acrodef{SA}    {speed augmentation}
\acrodef{SC}    {skip connection}
\acrodef{SepFormer}{separation transformer}
\acrodef{SiLU}  {sigmoid linear unit}
\acrodef{SISDR} {scale-invariant signal-to-distortion ratio}
\acrodef{SDR}   {signal-to-distortion ratio}
\acrodef{SNR}   {signal-to-noise ratio}
\acrodef{SRMR}  {speech-to-reverberation modulation energy ratio}
\acrodef{SSR}   {speech-to-speech ratio}
\acrodef{SP}    {signal processing}
\acrodef{STFT}  {short-time Fourier transform}
\acrodef{STOI}  {short-time objective intelligibility}
\acrodef{SOTA}  {state-of-the-art}
\acrodef{SW}    {shared weights}
\acrodef{ESTOI} {extended short-time objective intelligibility}
\acrodef{TasNet}{time-domain audio separation network}
\acrodef{TC}    {time-complexity}
\acrodef{TCN}   {temporal convolutional network}
\acrodef{TD-Conformer}{time domain conformer}
\acrodef{TSL}   {training signal length}
\acrodef{uPIT}  {utterance-level permutation invariant training}
\acrodef{UPGMA} {unweighted pair group method with arithmetic mean}
\acrodef{WER}   {word error rate}
\acrodef{WPE}   {weighted prediction error}
\acrodef{S} {small}
\acrodef{M} {medium}
\acrodef{L} {large}
\acrodef{XL} {extra-large}

\newcommand{\vek}[1]{\ensuremath{\mathbf{#1}}}    
\newcommand{\vekt}[1]{\ensuremath{\boldsymbol{\mathrm{#1}}}}

\newcommand{\Real}{\mathbb{R}}

\usepackage{cite} 
\usepackage[english]{babel}
\addto\extrasenglish{}
\addto\extrasenglish{}
\addto\extrasenglish{}
\addto\extrasenglish{}
\addto\extrasenglish{}
\addto\extrasenglish{}

\let\OLDthebibliography\thebibliography
\renewcommand\thebibliography[1]{
  \OLDthebibliography{#1}
  \setlength{\parskip}{0pt}
  \setlength{\itemsep}{0pt plus 0.5ex}
}

\title{On Time Domain Conformer Models for Monaural Speech Separation in Noisy Reverberant Acoustic Environments
\thanks{This work was supported by the Centre for Doctoral Training in Speech and Language Technologies (SLT) and their Applications funded by UK Research and Innovation [grant number EP/S023062/1]. This work was also funded in part by 3M Health Information Systems Inc.}}
\name{William Ravenscroft, Stefan Goetze, and Thomas Hain}
\address{Speech and Hearing Group, Dept.~of Computer Science, The University of Sheffield, Sheffield, UK}

\copyrightnotice{979-8-3503-0689-7/23/\$31.00~\copyright2023 IEEE}
\begin{document}
%
\maketitle
%


\begin{abstract}
Speech separation remains an important topic for multi-speaker technology researchers. 
\Acp{conformer} have performed well for many speech processing tasks but have been under-researched for speech separation. 
Most recent \ac{SOTA} separation models have been \acp{TasNet}.
A number of successful models have made use of \ac{DP} networks which sequentially process local and global information. 
\Acp{TD-Conformer} are an analogue of the \ac{DP} approach in that they also process local and global context sequentially but have a different time complexity function.
It is shown that for realistic shorter signal lengths, 
conformers are more efficient when controlling for feature dimension. 
Subsampling layers are proposed to further improve computational efficiency.
The best \ac{TD-Conformer} achieves $14.6$~dB and $21.2$~dB SISDR improvement on the WHAMR and WSJ0-2Mix benchmarks, respectively.
\end{abstract}
\noindent\textbf{Index Terms}: speech separation, conformer, speech enhancement, time domain, single channel

\section{Introduction}
Deep neural network models have led to significant improvements in monaural speech separation technology in recent years \cite{upit,tasnet,wdtcn,QDPN}. While impressive results have been attained on clean speech mixtures, noisy and reverberant mixtures still remain a challenging and active area of research \cite{FFASRHaebUmbach, MSA+13, WHAMR, atttasnet}.

Transformer \ac{TasNet} models have demonstrated \ac{SOTA} performance on numerous benchmarks in recent years \cite{dptnet, sepformer,tdanet,QDPN} 
due to their ability to process the global context of sequences. Many of these models used \ac{DP} networks \cite{dprnn, dptnet, sepformer}.
The conformer is a similar concept to \ac{DP} and quasi-\ac{DP} approaches \cite{dprnn,sepformer,QDPN}
but uses a single convolutional layer to process the local context instead of a more computationally complex \ac{RNN} or transformer layer. A benefit of this is reduced computational complexity as convolutional operations are more parallelizable along all dimensions and have linear \ac{TC} \cite{atttasnet}. A \ac{STFT}-based conformer model has been proposed for continuous speech separation in~\cite{cssconformer} which shows reasonable performance on the LibriCSS dataset but does not compare to popular time-domain techniques on other popular benchmarks as it is shown in \autoref{sec:results:final}, most likely due to its lower temporal resolution as \cite{IWAENCbestpaper} similarly demonstrated. 
A time-domain conformer and a \ac{TCN}-augmented conformer model were proposed for speech extraction in~\cite{sinhaconformer}.  The \ac{TCN} model performs best but has a much wider local context window, or \ac{RF}, than the pure conformer model. 
Questions remain about whether the full \ac{TCN} approach is necessary if a convolutional module in a conformer were to have a sufficiently wide kernel size~\cite{sinhaconformer, QDPN} and comparable model size, as this was not analysed in \cite{sinhaconformer}.

In this work, a single channel \ac{TD-Conformer} model is evaluated across different model sizes and computational expenditures from both theoretical and experimental perspectives. 
While \acp{TasNet} and conformer models are well studied, the combination of the two for separation tasks with a corresponding optimisation and performance analysis is as yet missing from the speech separation literature. 
Furthermore, this work demonstrates why it is an oversight in the area of speech separation research to not explore this particular combination in greater depth given the proposed model configuration in this paper is able to achieve close to \ac{SOTA} results with useful trade-offs in computational expenditure for the separation of shorter speech utterances.
It is shown in this work that, in the local-layer global-layer paradigm often used in speech separation \ac{DNN} models \cite{dprnn,dptnet,sepformer}, \ac{conformer} layers can be a more computationally suitable option in terms of efficiency.
In this work, a subsampling method is introduced which further reduces the computational \ac{TC} of the dot product attention in the transformer layers, similar to the SE-Conformer model \cite{se_conformer}. A number of evaluations are performed to assess the optimal \ac{RF} \cite{rfield} of the convolutional component in the conformer, the impact of subsampling on performance and computational complexity, and the benefits of the time domain approach over the \ac{STFT} model proposed in \cite{cssconformer}. Results are compared to a number of other \ac{SOTA} models in terms of performance, model size, and computational complexity across multiple benchmarks and acoustic conditions.

The remaining paper proceeds as follows. \autoref{sec:sigmod} introduces the signal model and \autoref{sec:td-conformer} the proposed \ac{TD-Conformer} model. Datasets and training configurations are explained in \autoref{sec:expsetup} and results are presented in \autoref{sec:results}. \autoref{sec:conclusions} gives final conclusions.

\section{Signal Model}\label{sec:sigmod}
A noisy discrete-time reverberant mixture signal $x[i]$ of $C$ speakers $s_c[i]$ with additive noise $n[i]$ at the microphone is defined as
\begin{equation}
    \label{eq:InputMixture}    x[i]=\sum_{c=1}^Cs_c[i]\ast h_c[i]+n[i],
\end{equation}
where the operator $\ast$ denotes the convolution and $h_c[i]$ is the \ac{RIR} corresponding to speaker $c \in\{1,\ldots,C\}$. In this work the aim is to find an estimate for each of the $C$ speech signals, denoted by $\hat{s}_c[i]$.

\section{TD-Conformer Separation Networks} \label{sec:td-conformer}
\begin{figure}[!h]
    \centering
    \includegraphics[width=1.0\columnwidth]{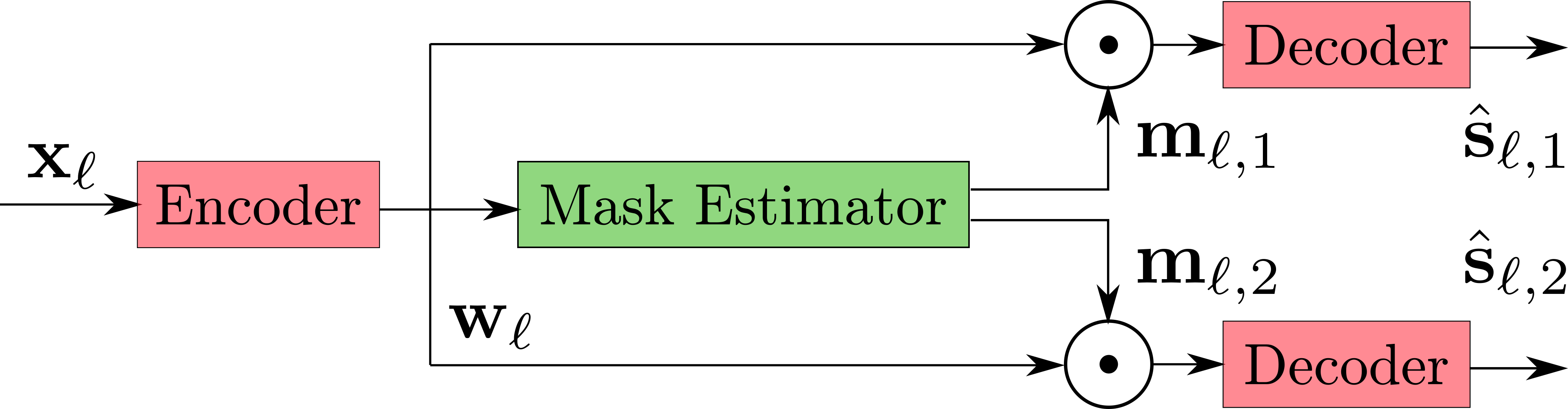}
    \caption{\Ac{TD-Conformer} \ac{TasNet} model diagram, exemplary for $C=2$ speakers for each frame index $\ell$. Operator $\odot$ denotes the Hadamard product and $\mathbf{w}_\ell$ the vector of encoded features for frame $\ell$.}
    \label{fig:contasnet}
\end{figure}

The proposed \ac{TD-Conformer}, described in the following, is a \ac{TasNet} composed of three main components: a feature encoder, a mask estimation network and a decoder as depicted in \autoref{fig:contasnet}. The encoder is a learnable filterbank \cite{tasnet, Ditter} and the decoder performs the inverse function to convert encoded features back into the time domain. The mask estimation network calculates $C$ masks $\mathbf{m}_{\ell,c}$ for each time frame $\ell$ to obtain estimates $\mathbf{\hat{s}}_{\ell,c}$ of the clean input speech signals $\mathbf{s}_{\ell,c}$. Boldface letters indicate vectors capturing respective frames of samples of the quantities in \autoref{sec:sigmod}.

\subsection{Encoder}
Similar to \cite{sepformer, convtasnet}, the mixture signal $x[i]$ in (\ref{eq:InputMixture}) is segmented into $L_\vek{x}$ overlapping blocks $\vek{x}_\ell$ of size $L_\mathrm{BL}$ which are encoded into $\vek{w}_\ell\in\Real ^{1\times N}$ using a 1D convolutional layer with weights $\vek{B} \in \Real^{L_\mathrm{BL}\times N}$ for $N$ output channels, followed by a \ac{ReLU} activation function $\mathcal{H}_\mathrm{enc} : \Real^{1 \times N} \rightarrow \Real^{1 \times N}$ producing encoded features
\begin{equation}
\vek{w}_\ell=\mathcal{H}_\mathrm{enc}\left(\vek{x}_\ell \vek{B}\right) \in \Real^{L_\mathrm{x} \times N}.
\end{equation}

\subsection{Conformer Mask Estimation Network}
\begin{figure}[!h]
    \centering
    \includegraphics[width=\columnwidth]{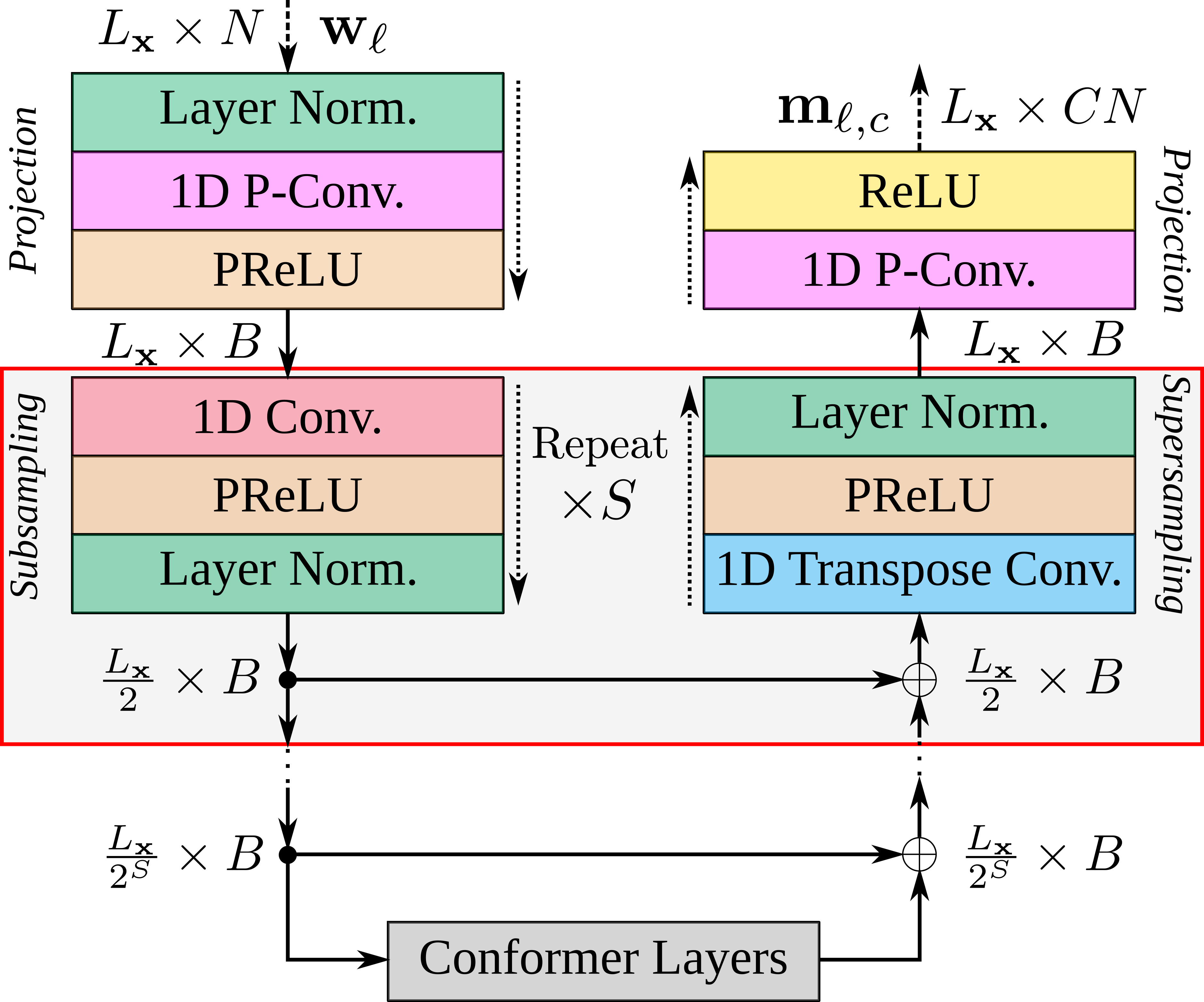}
    \caption{Proposed TD-Conformer mask estimation network structure with subsampling and supersampling layers to reduce and increase the temporal resolution, and also enabling a reduction of the \acf{TC} in the conformer layers.
    }
    \label{fig:masknet}
\end{figure}

The mask estimation network is based on the conformer architectures proposed in \cite{conformer,se_conformer}. A diagram of the conformer mask estimation network is shown in \autoref{fig:masknet}.

The input sequence of features $\vek{w}_\ell$ is 
normalized using layer normalization \cite{ba2016layer} before being projected from dimension size $N$ to $B$ using a \ac{P-Conv} layer followed by a \ac{PReLU} activation. This results in a sequence of features of shape $L_\vek{x}\times B$. This sequence is then fed through $S$ subsampling layers each of which is a 1D convolutional layer of a fixed kernel size of $4$ and stride of $2$ thus reducing the temporal dimension by a factor of $2$ giving a sequence of shape $\frac{L_\vek{x}}{2^S} \times B$. Each subsampling layer has a skip connection to a respective supersampling block composed of a 1D transposed convolutional layer, \ac{PReLU} and layer normalization. This structure increases the temporal dimension by a factor of $2$ to restore the sequence length to $L_\vek{x}$. In between the subsampling and supersampling blocks are a series of $R$ conformer layers.  

\begin{figure}[!t]
    \centering
    \includegraphics[width=\columnwidth]{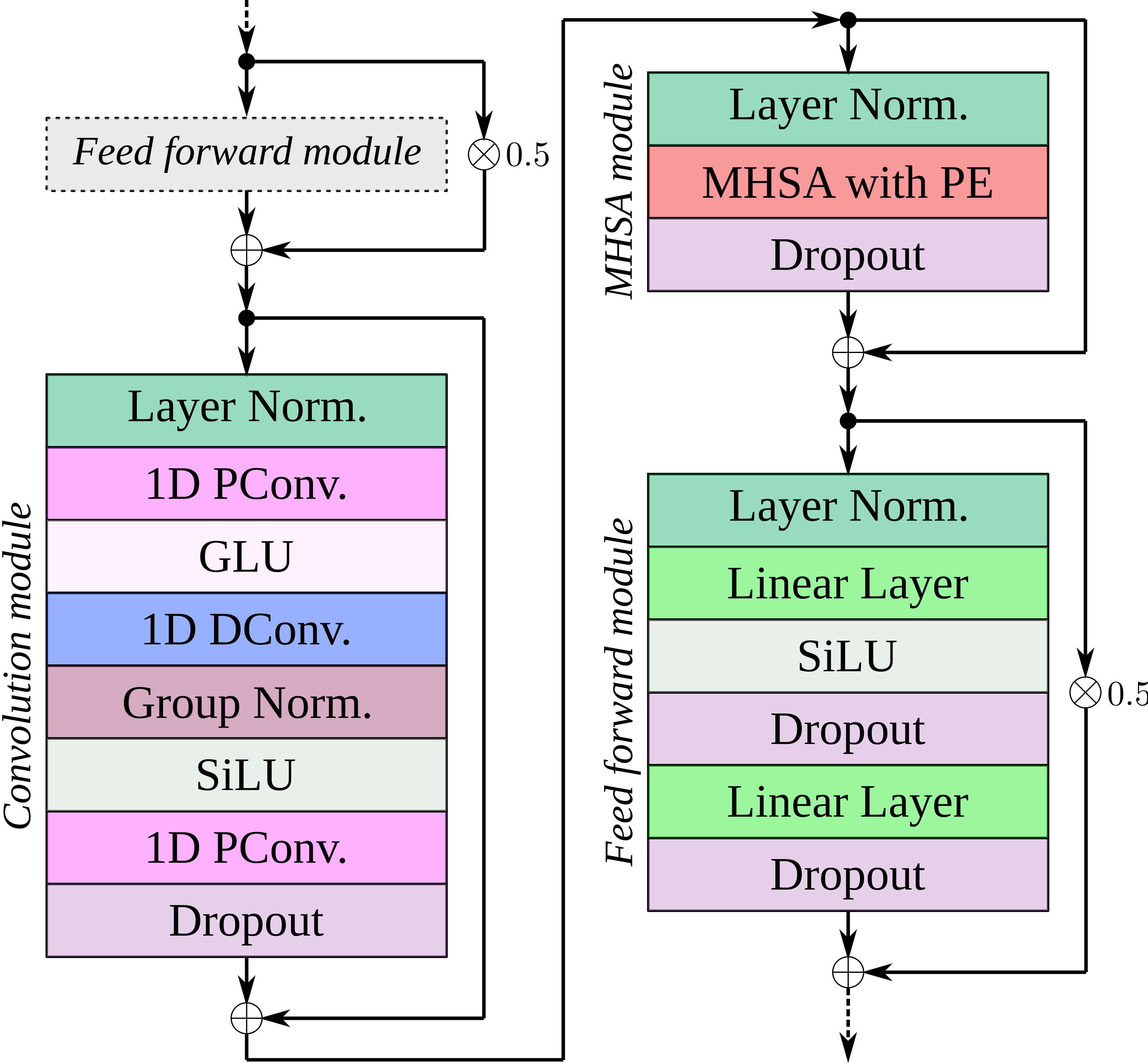}
    \caption{Diagram of a single Conformer layer composed of feed-forward, convolution and \ac{MHSA} modules. Note the first feed-forward module is identical to the final module but its details are omitted for brevity.}
    \label{fig:conformer}
\end{figure}

The conformer layers are shown in detail in \autoref{fig:conformer} and are composed of a feed-forward module, a convolution module, a \ac{MHSA} module and another feed-forward module. 
The convolution module comes before the \ac{MHSA} module contrasting the original conformer~\cite{conformer} which has \ac{MHSA} first. This is so the model processes the local context first similar to the \ac{DP} models proposed in~\cite{dprnn, sepformer}.
The two feed-forward modules are composed of layer normalization, a linear layer with \ac{SiLU} activation~\cite{silu}, dropout \cite{dropout} and then another linear layer followed by another dropout. Each feed-forward module has a weighted residual connection from input to output.
The convolution module is composed of layer normalization, a \ac{P-Conv} with \ac{GLU} activation, a \ac{D-Conv} with kernel size $P$, group normalization \cite{groupnorm}, a \ac{SiLU} activation and lastly a \ac{P-Conv} layer followed by dropout. For the group normalization, the number of groups is equal to the number of input channels. A residual connection goes from the input to the output of the module.
The \ac{MHSA} module is composed of layer normalization and \ac{MHSA} with relative \ac{PE}~\cite{Vaswani} followed by dropout. The \ac{MHSA} module has a residual connection around the entire module.

\subsection{Conformers vs. Dual-Path Transformers}\label{sec:td-conformer:vsdpt}
\begin{figure}[!ht]
    \centering
    \includegraphics{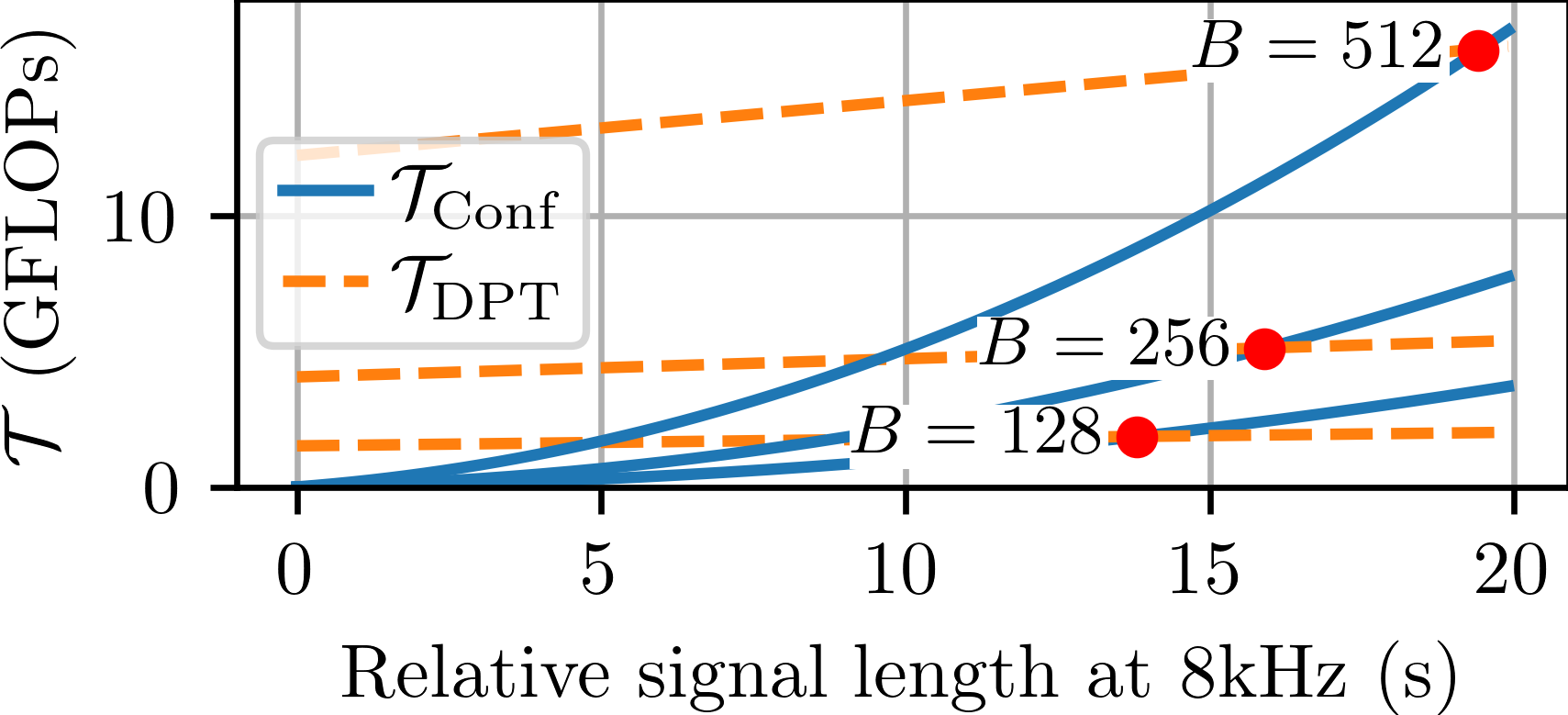}
    \caption{Comparison of \acp{TC} measured in {GFLOPS} for a conformer layer and a \ac{DP} Transformer layer for different feature dimensions $B\in \{128,256,512\}$ over relative signal length in seconds. Note that $S=0$ for the Conformer layer here and $P=P'=250$ is used as it is equal to the best performing configuration of the \ac{DP} model in \cite{sepformer}.}
    \label{fig:tc_comp}
\end{figure}

The conformer layers are proposed in analogy to the \ac{DP} transformer layers in the widely researched SepFormer model~\cite{sepformer,IWAENCbestpaper}. Note that other \ac{DP} transformer layers have been proposed, such as in \cite{dptnet}, but we disregard these here as they are more computationally expensive and less performant than SepFormer \cite{sepformer,sudormrf}. 
In this section, the respective \ac{TC} functions of the conformer layer and the \ac{DP} transformer layer are modelled to demonstrate that under certain (often more realistic) signal lengths and feature dimensions, conformers are less computationally complex than \ac{DP} transformers.
The intra-transformer is compared to the convolutional module of the conformer as a local context layer and the inter-transformer is compared to \ac{MHSA} modules of the conformer as a global context layer. The \ac{TC} for a conformer layer is defined as 
\begin{equation}\label{eq:conf_tc}
    \mathcal{T}_\mathrm{Conf} = \frac{L_\vek{x}}{2^S} \left(PB+B^2\right)+\frac{L_\vek{x}^2}{2^{2S}}B+B^2\frac{L_\vek{x}}{2^S}. 
\end{equation}
The \ac{TC} for a dual-path transformer layer in the style of SepFormer~\cite{sepformer} is defined as
\begin{equation}
    \mathcal{T}_\mathrm{DPT} = \left(\frac{L_\vek{x}}{2P'}+\frac{P'}{2}\right) \left(P'^2B+B^2P'\right) + \frac{L_\vek{x}}{4P'^2}B+B^2\frac{L_\vek{x}}{2P'},
\end{equation}
where $P'$ represents the chunk size \cite{sepformer}.
The time-com\-plexities for \ac{DP} transformer and conformer layers with a feature sequence from an 8kHz piece of audio encoded with block length $L_\mathrm{BL}=16$ are shown in \autoref{fig:tc_comp} for different feature dimensions $B$. 
The average and maximum signal lengths in the WHAMR \textit{tt} evaluation set used later in \autoref{sec:results}, are $5.79$s and $13.87$s, respectively \cite{tsllimits}. Hence, in evaluations on this dataset, the conformer layer is the less computationally complex option overall. For larger $B$ values the conformer more likely has lower \ac{TC} than the \ac{DP} transformer. Another benefit of the conformer is that, assuming a small number of subsampling layers, it has a much higher temporal resolution than the \ac{DP} approach when processing the global context.
\begin{figure}[!ht]
    \centering
    \includegraphics{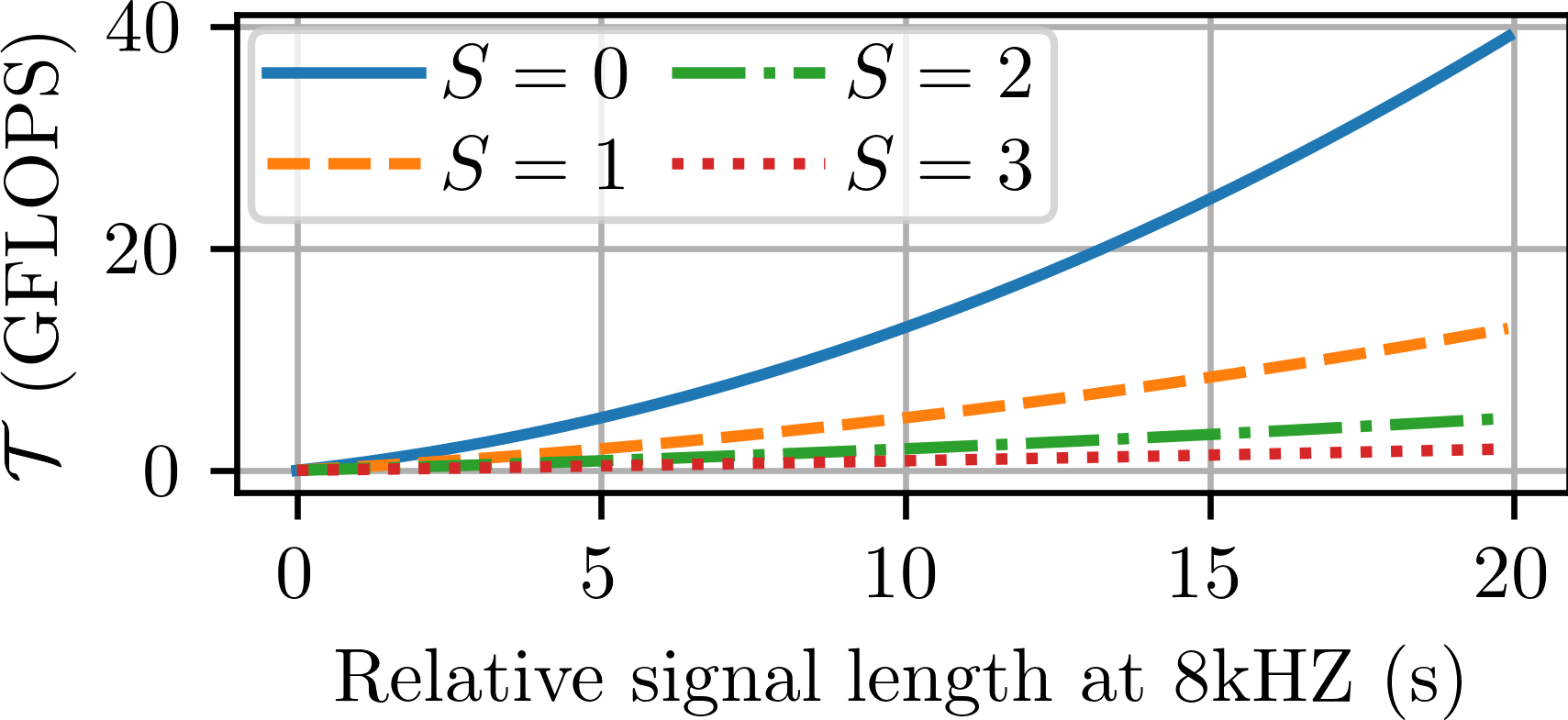}
    \caption{Comparison of conformer \ac{TC} function over relative signal length for varying subsampling layers $S\in\{0,1,2,3\}$.}
    \label{fig:ss_complexity}
\end{figure}
\autoref{fig:ss_complexity} demonstrates that every additional subsampling layer can significantly reduce the \ac{TC} of the conformer layer.
The impact of increasing the subsampling on overall performance is explored in more depth later in \autoref{sec:ss_layers}. Note that the \ac{DP} transformer topology also has its own implicit subsampling strategy akin to a strided view of the output tensor from the local transformer layer which reduces the computational complexity but also the temporal resolution significantly \cite{QDPN,sepformer}.

\subsection{Objective function}
An \ac{SISDR} objective function \cite{LeRoux,tasnet} with an \ac{uPIT} wrapper \cite{upit} is used to train all of the models. The \ac{SISDR} loss function is defined as
\begin{align} 
\label{eq:DefSISDR}
\mathcal{L}(\hat{\vekt{s}},\vekt{s})&: 
= \frac{1}{C}\sum_{c=1}^C- 10\log_{10} \frac{\left\Vert \frac{\langle \hat{\vekt{s}}_c,\vekt{s}_{c}\rangle \vekt{s}_{c}}{\Vert \vekt{s}_{c}\Vert^{2}}
\right\Vert^{2}}{\left\Vert\hat{\vekt{s}}_c-\frac{\langle \hat{\vekt{s}}_c,\vekt{s}_{c}\rangle \vekt{s}_{c}}{\Vert \vekt{s}_{c}\Vert^{2}}\right\Vert^{2}}.
\end{align}

\section{Experimental Setup}\label{sec:expsetup}
\subsection{Data}
The WHAMR and WSJ0-2Mix datasets are used for analysing the proposed TD-Conformer models \cite{Isik,WHAMR}. WSJ0-2Mix is a 2-speaker mixture corpus~\cite{Isik}. Speech segments from the WSJ0 corpus are mixed at \acp{SSR} between $0$ to $5$~dB.
WHAMR is a noisy reverberant extension of WSJ0-2Mix. Ambient recorded noise is mixed with the speakers at \acp{SNR} between $-6$ and $3$~dB. Simulated \acp{RIR} are used to reverberate the speech mixtures with reverberation time T60 values ranging between $0.1$s and $1$s.

\subsection{Network configurations}\label{sec:netconfig}
Four model configurations are proposed to vary the model size in increasing internal mask estimation feature dimension $B$ of $128$ (denoted as \ac{S}), $256$ (\ac{M}), $512$ (\ac{L}) and $1024$ (\ac{XL}). The encoder has a fixed output dimension of $N=256$ and a kernel size of $L_\mathrm{BL}=16$ with a $50\%$ stride of $8$ samples. The number of conformer layers is fixed to $R=8$, the same as the number of \ac{DP} layers in~\cite{sepformer}. In the results sections the number of subsampling layers $S$ and conformer kernel size $P$ are modified to gain a better understanding of their impact on separation performance and computational cost. For all evaluations a dropout of $10\%$ is used.

\subsection{Training configurations}
All models are trained using an initial learning rate of $5\times 10^{-5}$. Learning rates are fixed for the first $90$ epochs and then halved if there is no performance improvement after $3$ epochs. Training is performed over $200$ epochs. Training examples were limited to $4$ seconds. In \cite{tsllimits} it was shown that this signal length is in an optimal range to reduce training time without impacting overall performance for the datasets used in this paper. By limiting the \acp{TSL}, it enables the use of a batch size of $4$ even with the largest XL models proposed in \autoref{sec:netconfig}. This contrasts the best performing SepFormer model where it has been shown that even with comparable \ac{TSL} limits the largest batch size it was possible to use was $2$ \cite{tsllimits} on the same GPU used in this paper, a 32GB Nvidia V100. Further to the discussion in \autoref{sec:td-conformer:vsdpt}, the reason for this difference is that despite both the TD-Conformer XL model and the Sepformer using an internal feature dimension of $1024$, the use of \ac{MHSA} in the Sepformer for processing local information consumes a lot more memory for an utterance of $4$s in length.
An open-source SpeechBrain \cite{speechbrain} \emph{recipe} is provided with this work to enable other researchers to reproduce the results in this paper.\footnote{GitHub link to SpeechBrain conformer recipe: \url{https://github.com/jwr1995/PubSep}.}

\subsection{Evaluation Metrics}
The main separation evaluation metric used is \ac{SISDR} improvement over the noisy mixture, denoted by $\Delta$ \ac{SISDR} and measured in~dB. Where relevant, computation expenditure is report using \ac{MACs}. \Ac{MACs} are calculated using the \textit{thop} \cite{thop} toolkit on a signal of $5.79$s in length, equal to the mean signal length in the WHAMR test set. This is to be more reflective of a realistic input as the TD-Conformer models contain quadratic time complexities. Model size is measured in parameter count where relevant.

\section{Results}\label{sec:results}

\subsection{Varying kernel sizes}\label{sec:ks}
\begin{figure}[!ht]
    \centering
    \includegraphics{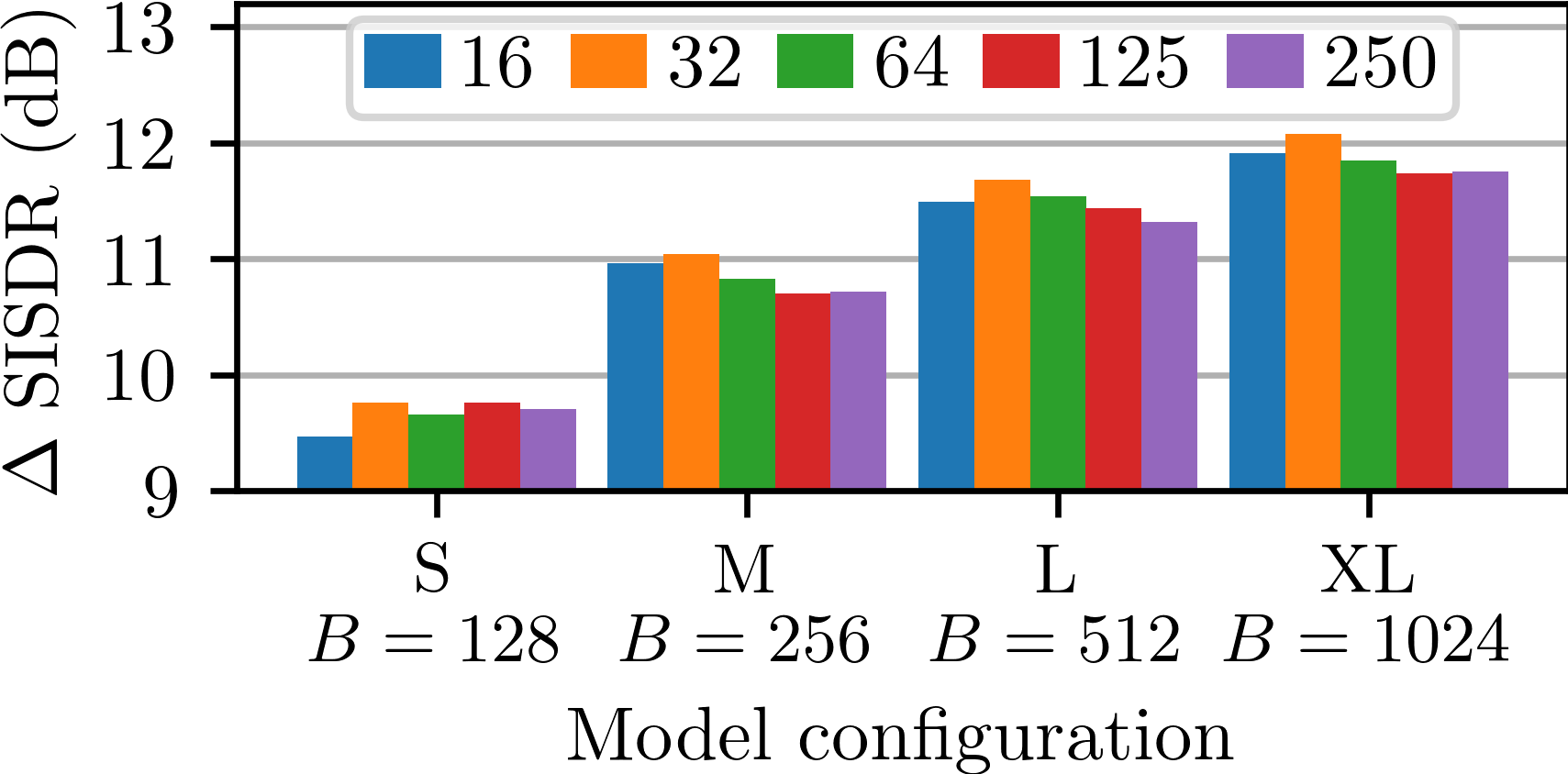}
    \caption{Performance in $\Delta$ SISDR for conformer layer kernel sizes $P \in \{16, 32, 64, 125, 250\}$ and different model sizes based on $B$.}
    \label{fig:ks}
\end{figure}

In this section, the kernel size $P$ of the \ac{D-Conv} layer in the conformer layer is varied for different feature dimensions $B$ to evaluate if there exists a common $P$ value.
Five $P$ values are evaluated: $\{16, 32, 64, 125, 250\}$. The values $16$ and $32$ are similar to the kernel sizes in the original conformer model \cite{conformer} as well as in~\cite{cssconformer, se_conformer}. The values $64$, $125$ and $250$ were selected to give the \ac{D-Conv} layer a comparable \ac{RF} to the local transformer layer of popular \ac{DP} models such as SepFormer \cite{sepformer} where the chunk size is set to $250$ for the equivalent $L_\mathrm{BL}=16$ model configuration.
The results in \autoref{fig:ks} show that for all models the performance in $\Delta$~SISDR peaks for a kernel size of $P=32$.
For sampling rate $\mathrm{f}_s$, the relative \ac{RF} of each convolution module in seconds is defined as
\begin{equation}\label{eq:RF}
    \mathcal{R}_\mathrm{conv}(S,P, L_\mathrm{BL}, \mathrm{f}_s)=\frac{1}{\mathrm{f}_s}\left({2^{S-1}L_\mathrm{BL}}P+\frac{L_\mathrm{BL}}{2}\right)
\end{equation}
which for the configuration $\{S,P, L_\mathrm{BL}, \mathrm{f}_s\}=\{2,32,16,$ $8000\}$ is $0.129$s. This \ac{RF} value is used later in \autoref{sec:results:final} for optimising the model hyperparameters.

\subsection{Varying the number of subsampling layers}\label{sec:ss_layers}
Altering the number of subsampling layers changes the temporal resolution of the input encoded features to the Conformer layers in the mask estimation network. It also inversely affects the overall model size, i.e.~more subsampling layers result in lower temporal resolution but slightly larger model size. In this second experiment 
the number of subsampling layers $S$ is varied from $0$ to $3$. Note that for $S=0$, a smaller batch size of $2$ has to be used due to the increased memory consumption of the transformer layers. A fixed kernel size of $P=32$ is used.
\begin{figure}[!ht]
    \centering
    \includegraphics{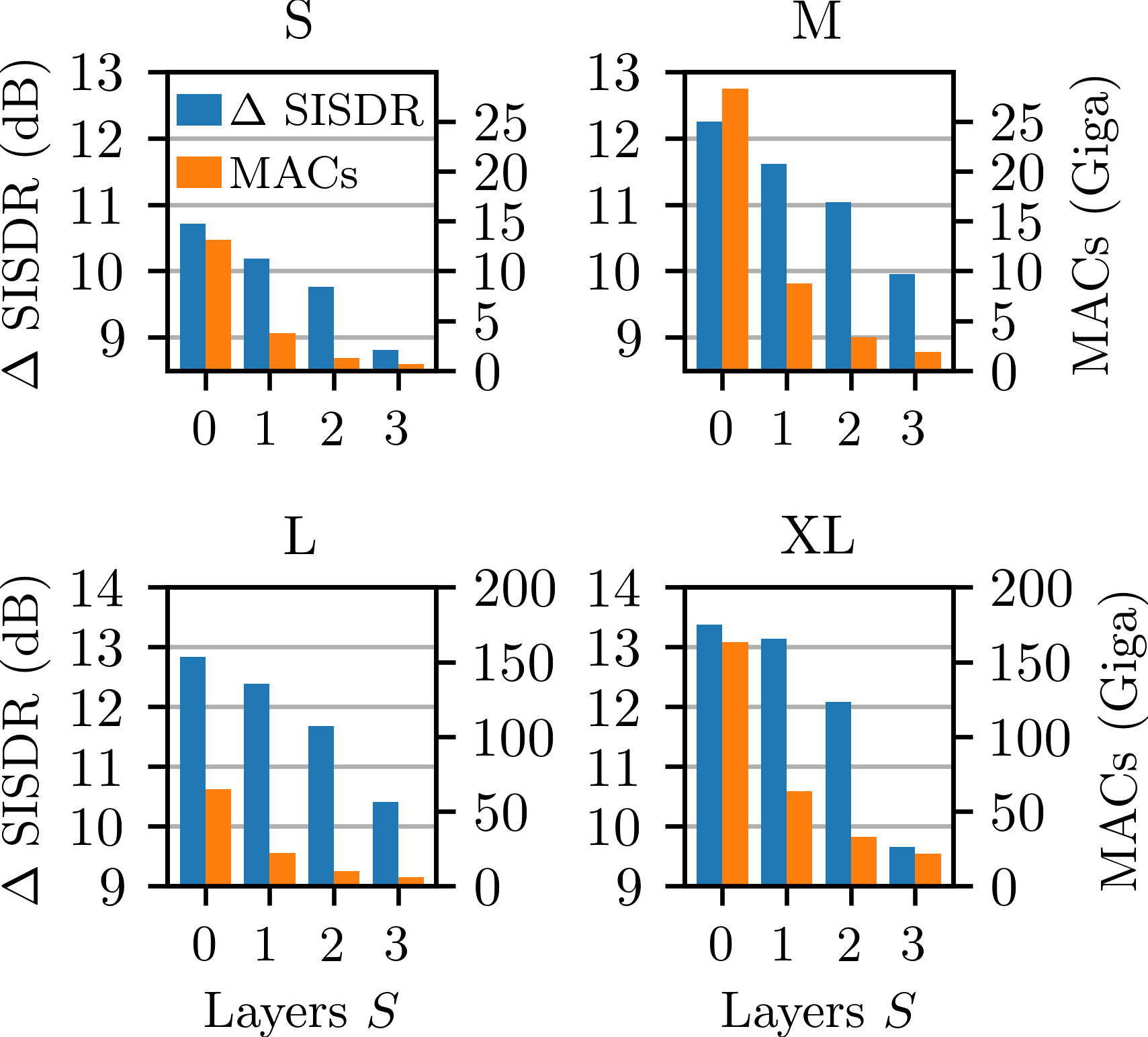}
    \caption{Performance over number of  subsampling layers $S$ for all conformer model sizes (S, M, L \& XL) with respective computational cost in \ac{MACs} exemplary for signal of length $5.79$s.}
    \label{fig:ss}
\end{figure}
From \autoref{fig:ss} it is notable that the difference in performance between $S=0$ and $S=1$ layers is less than between $S=1$ and $S=2$ and likewise then for $S=2$ and $S=3$. It is also noticeable that for the smaller model sizes the reduction in performance for each additional subsampling layer is also smaller.
The $S=0$ configuration gives the best overall performance for both the S and XL TD-conformer models. This is expected as $S=0$ gives the \ac{MHSA} layers in the conformer the highest temporal resolution when processing the global context. This improvement comes at a significant computational cost however as can be seen from the \ac{MACs} reported in \autoref{fig:ss} (note the scales vary for each row). This provides some justification for using a small number of subsampling layers at the benefit of significant reductions in computational requirements. 

\subsection{Hyper-parameter optimization and comparison to other models}\label{sec:results:final}
In the following, TD-Conformer models are compared with several other discriminative supervised speech separation models. To find optimal configurations for the conformer models, a set of TD-Conformer models with $S=1$ subsampling layers and kernel sizes $P\in\{64,125\}$ are trained. Models with no subsampling  ($S=0$) are not evaluated, due to unreasonably long training times on the hardware available for only minor improvements in performance (cf.~\autoref{sec:ss_layers}). The kernel size $P=64$ was specifically chosen as it gives an equal \ac{RF} to the optimal configuration $\{P,S\}=\{32,2\}$ in  \autoref{sec:ks}, cf.~(\ref{eq:RF}). \Ac{DM}, as implemented in \cite{wavesplit}, is also used to maximise model performance for each of the models. Results are reported with and without \ac{DM}.
 \Ac{MACs} are reported for all models for which an open source implementation was available for evaluation \cite{convtasnet,sudormrf,deformtcn,sepformer}. 
Performance in terms of $\Delta$~SISDR for the WSJ0-2Mix anechoic speech mixture dataset \cite{Isik} is shown for completeness.

\autoref{tab:final_table} shows that all TD-Conformer models outperform the \ac{STFT} conformer model proposed in \cite{cssconformer}. The small TD-Conformer-S models outperform the similarly sized and complex DTCN and SuDoRMRF++ baselines \cite{deformtcn,sudormrf} with \ac{DM} on WSJ0-2Mix and show comparable performance on WHAMR without \ac{DM}. The medium TD-Conformer-M models give comparable performance to the similarly sized SkiM baseline \cite{skim} on the WSJ0-2Mix benchmark but with less than half the number of \ac{MACs}. The large TD-Conformer-L models give better performance than the SepFormer baseline on the WHAMR benchmark and comparable performance on WSJ0-2Mix with a similar model size but roughly a third of \ac{MACs}. The TD-Conformer-XL model outperforms the much larger \ac{QDPN} model on WHAMR with the best model giving $14.6$~dB $\Delta$~SISDR. The TD-Conformer does not quite reach parity with the more recent MossFormer-L and TF-GridNet \cite{mossformer, tfgridnettasl} models however the largest TF-GridNet model has significantly higher computational expenditure and MossFormer is an augmented version of a Conformer model. Thus the results here help to validate the design choice of this approach.
\begin{table}[!ht]
\setlength\tabcolsep{1.5pt}
\resizebox{\columnwidth}{!}{%
\begin{tabular}{|c|c|c|c|c|c|}
\hline
\rowcolor[HTML]{C0C0C0} 
\textbf{}                        & \textbf{}    & \multicolumn{2}{c|}{\cellcolor[HTML]{C0C0C0}\textbf{$\Delta$ SISDR (dB)}}                             & \textbf{}     & \textbf{}       \\ \cline{3-4}
\rowcolor[HTML]{C0C0C0} 
\textbf{Model}                   & \textbf{$P$} & \multicolumn{1}{c|}{\cellcolor[HTML]{C0C0C0}\textbf{W-2Mix}} & \cellcolor[HTML]{C0C0C0}\textbf{WHAMR} & \textbf{MACs} & \textbf{Params} \\ \hline
Conv-TasNet \cite{convtasnet}    & -          &  15.6                      & 9.7*& 3.6G           & 5.1M               \\
STFT-Conformer \cite{cssconformer}      &        -      &    10.8*      &    6.7*        &           \textbf{1.8G}     &   57.5M          \\
SuDoRMRF++ +DM \cite{sudormrf} & -          & 17                      &      -                           & 2.7G          & 2.7M               \\ 
DTCN \cite{deformtcn}            & -          & 15.6                    & 10.2                              & 3.7           & 3.6M               \\
DTCN+DM \cite{deformtcn}         & -          & 17.2                    & 11.1                              & 3.7G           & 3.6M               \\
SkiM \cite{skim}                 & -          & 18.3                    & -                                 & 19.7G          & 5.9M               \\
SepFormer \cite{sepformer}       & -          & 20.4                    &  11.5*                             & 60.7G        & 25.6M                \\
SepFormer+DM \cite{sepformer}    & -          & 22.3                    & 14                                & 60.7G        & 25.6M                \\
QDPN \cite{QDPN}                 & -          & 22.1                    & 13.1                              & -             & 200M               \\
QDPN+DM \cite{QDPN}              & -          & \textbf{23.6}                    & 14.4                              & -             & 200M               \\ 
MossFormer-L+DM \cite{mossformer} & - & 22.8 & 16.3 & 42.8G & 42.1M \\
TF-GridNet (Tab.  XIII) \cite{tfgridnettasl}& - & 22.0 & - & 29.8G & 6.8M \\
TF-GridNet \cite{tfgridneticassp,tfgridnettasl}& - & 23.4 & \textbf{17.1} & 228.2G & 14.3M \\\hline
TD-Conformer-S                   & 64  & 15.8         & 10.5             & 3.7G   & \textbf{1.8M}    \\
TD-Conformer-S                   & 125 & 15.9         & 10.5             & 3.7G   & \textbf{1.8M }   \\
TD-Conformer-S+DM                & 64  & 17.4         & 9.7              & 3.7G   & \textbf{1.8M}    \\
TD-Conformer-S+DM                & 125 & 17.5         & 9.7              & 3.7G   & \textbf{1.8M}    \\ \hline
TD-Conformer-M                   & 64  & 17.7         & 11.7             & 8.5G   & 6.7M    \\
TD-Conformer-M                   & 125 & 17.8         & 11.6             & 8.6G   & 6.8M    \\
TD-Conformer-M+DM                & 64  & 18.1         & 12.0             & 8.5G   & 6.7M    \\
TD-Conformer-M+DM                & 125 & 18.8         & 11.9             & 8.6G   & 6.8M    \\ \hline
TD-Conformer-L                   & 64  & 19.5         & 12.3             & 21.9G  & 25.9M   \\
TD-Conformer-L                   & 125 & 19.7         & 12.5             & 22.0G  & 26.2M   \\
TD-Conformer-L+DM                & 64  & 20.3         & 13.4             & 21.9G  & 25.9M   \\
TD-Conformer-L+DM                & 125 & 20.2         & 13.2             & 22.0G  & 26.2M   \\ \hline
TD-Conformer-XL                  & 64  & 20.4         & 13.1             & 63.6G  & 102.2M  \\
TD-Conformer-XL                  & 125 & 20.3         & 13.0             & 63.9G  & 102.7M  \\
TD-Conformer-XL+DM               & 64  & 21.1         & {14.6}             & 63.6G  & 102.2M  \\
TD-Conformer-XL+DM               & 125 & 21.2         & 14.3             & 63.9G  & 102.7M  \\ \hline
\end{tabular}%
}
\caption{$\Delta$ SISDR results for various TD-Conformer models with $S=1$ compared to other separation models on the WSJ0-2Mix (abbrev. W-2Mix) and WHAMR benchmarks.  *~indicates results not included in the respective paper cited for a model. }
\label{tab:final_table}
\end{table}


\section{Conclusion}\label{sec:conclusions}
In this paper, a single-channel time-domain conformer speech separation model was introduced and evaluated. It was shown to reach comparable $\Delta$~SISDR performance as \ac{SOTA} models on the WHAMR and WSJ0-2Mix benchmarks.
Using conformer layers in place of \ac{DP} transformer layers was demonstrated to reduce the \ac{TC} of processing local information whilst increasing the \ac{TC} for processing global context. A benefit of increased global \ac{TC} is that it gives the global context layer a higher temporal resolution as demonstrated by varying the number of subsampling layers before the conformer layers in the proposed network. The proposed TD-Conformer-XL model achieves $14.6$~dB $\Delta$ \ac{SISDR} on the WHAMR benchmark. The smallest TD-Conformer-S model outperforms a number of larger and similarly complex models on the WSJ0-2Mix and WHAMR benchmarks.

\bibliographystyle{IEEEtran}
\bibliography{refs}

\end{document}